\documentclass[twocolumn,showpacs,preprintnumbers,amsmath,amssymb]{revtex4}
\usepackage{graphicx}% Include figure files
\usepackage{dcolumn}% Align table columns on decimal point
\usepackage{bm}% bold math

\begin{document}

\preprint{APS/123-QED}

\title{The Effect of Dynamical Parton Recombination on Event-by-Event Observables\\}

\author{Stephane Haussler}
\email{haussler@fias.uni-frankfurt.de}
\affiliation{%
Frankfurt Institute for Advanced Studies (FIAS), Max-von-Laue-Str.~1, D-60438 Frankfurt am Main, Germany
}%

\author{Stefan Scherer}%
\affiliation{%
Frankfurt Institute for Advanced Studies (FIAS), Max-von-Laue-Str.~1, D-60438 Frankfurt am Main, Germany
}%

\author{Marcus Bleicher}%
\affiliation{%
Institut f\"ur Theoretische Physik, Johann Wolfgang Goethe-Universit\"at, Max-von-Laue-Str.~1, D-60438 Frankfurt am Main, Germany
}%

\date{\today}% It is always \today, today,

\begin{abstract} 
Within a dynamical quark recombination model we explore various proposed 
event-by-event observables sensitive to the microscopic structure of the 
QCD-matter created at RHIC energies. Charge fluctuations, charge transfer fluctuations and
baryon-strangeness correlations are computed from a sample of central
Au+Au events at the highest RHIC energy available ($\sqrt{s_{NN}}$=200~GeV). 
We find that for all explored observables, the calculations yield the values predicted 
for a quark-gluon plasma only at early times of the evolution, whereas the final 
state approaches the values expected for a hadronic gas. 
We argue that the recombination-like hadronization process itself
is responsible for the disappearance of the predicted deconfinement signatures. 
This might explain why no fluctuation signatures for the transition between quark and hadronic
matter was ever observed in the experimental data up to now. However, it might also be interpreted
as a clear indication for a recombination like hadronization process at RHIC.
\end{abstract}



\maketitle

It is widely believed that a phase transition from a quark-gluon
plasma (QGP) to hadronic matter occurs in central ultra-relativistic
heavy-ions collisions at RHIC. In order to study the properties of the
extremely heated and compressed matter created in these events,
numerous probes based on fluctuations have been proposed \cite{Gazdzicki:1992ri,Mrowczynski:1997kz,Bleicher:1998wd,Bleicher:1998wu,Stephanov:1998dy,Jeon:1999gr,Stephanov:1999zu,Mrowczynski:1999sf,Capella:1999uc,Mrowczynski:1999un,Bleicher:2000tr,Asakawa:2000wh,Bleicher:2000ek,Jeon:2001ka,Sa:2001ma,Koch:2001zn,Hatta:2003wn,Ferreiro:2003dw,Mekjian:2004qf,Mrowczynski:2004cg,Shi:2005rc,Jeon:2005kj,Koch:2005vg,Konchakovski:2005hq,Cunqueiro:2005hx,Torrieri:2005va,Armesto:2006bv}.
For a comprehensive overview in the physics of event-by-event fluctuations we refer the reader to \cite{Jeon:2003gk}.
Among them, especially charge ratio fluctuations, charge transfer fluctuations and
baryon-strangeness correlations were prominently proposed to pin down the formation of a 
deconfined phase at RHIC
\cite{Jeon:2000wg,Asakawa:2000wh,Jeon:2001ue,Zhang:2002dy,Pruneau:2002yf,Shi:2005rc,Jeon:2005kj,Koch:2005vg,Haussler:2005ei}. 

These observables are based on event-by-event fluctuations of conserved charges within a given
rapidity range and are sensitive to the microscopic nature of the
matter. It was pointed out that these quantities reflect the properties
of the system in the first instant of the collision and should survive the
whole course of the evolution of the system. The argument in favour of the 
survival of the signal through all the
stages of the collision for the above mentioned fluctuations probes is
the following: With a strong transverse and longitudinal flow, locally conserved
quantities (charge, baryon number and strangeness) will be frozen in a
given rapidity window because the expansion is too quick for the
charges to move out of the respective  rapidity slice. Thus, if a QGP is
created, the fluctuation of these quantities should survive further evolution 
through the hadronic phase. 

It is clear that the size of the rapidity window 
for the fluctuation study must not be too wide in order to
avoid global conservation which would lead to a vanishing signal, but
also neither too small to avoid purely statistical fluctuations and the transport of charges in and out of
the window by hadronic rescattering. The generally accepted rapidity width  is
of the order of $\Delta y = 0.5-1$ units in rapidity. In contrast to the RHIC energies explored here, 
it was argued that the diffusion rate
for secondaries at the CERN-SPS might be strong enough to blur the fluctuation signal
almost to the (observed) resonance gas value \cite{Shuryak:2000pd}.

A key point that is usually not addressed in the discussion of fluctuation signals 
is the influence of hadronization itself. A possible mechanism for the parton-hadron
transition is the recombination of quarks and anti-quarks into hadrons
\cite{Molnar:2003ff,Hwa:2003ic,Fries:2003vb,Fries:2003kq,Greco:2003xt}. Elliptic
flow and nuclear modification factor $R_{AA}$ measurements at RHIC
\cite{Adams:2003am,Adler:2003kt} have given strong evidence
supporting recombination as the mechanism responsible for
hadronization. A first exploratory study on the influence of parton recombination on charge fluctuation was 
performed in \cite{Nonaka:2005vr}. There it was shown that the
coalescence of quarks through the recombination mechanism does indeed lead to results
compatible with the available experimental data on charge
fluctuations at RHIC.

In this paper, we study charge ratio fluctuations, charge transfer
fluctuations and baryon-strangeness correlations with a dynamical
recombination model (the quark Molecular Dynamics model, qMD
\cite{Hofmann:1999jx,Scherer:2001ap,Scherer:2005sr} ). To pin down the influence
of the hadronization process in detail we explore the suggested quantities 
over the whole time evolution of the system from the pure quark stage to the final 
hadrons. The set of events consist of central Au+Au collisions
at the highest RHIC energy available ($\sqrt{s_{NN}}=200$~GeV).
We will finally conclude that the hadronization process itself is responsible
for the change of all investigated observables from the initially partonic value 
to the finally observable hadronic value. Thus, providing evidence for a recombination like
hadronization mechanism at RHIC energies.

The qMD model \cite{Hofmann:1999jx,Hofmann:1999jy,Scherer:2001ap,Scherer:2005sr} employed here is a 
semi-classical molecular dynamics approach 
where quarks are treated as point-like particles carrying color charges and interact via a linear
heavy quark potential. Initial conditions\footnote{It should be noted that the qualitative results of the present study 
are not restricted to any specific initial state. The UrQMD model is solely used to provide an exemplary 
initial state after the initial $q\overline q$ production has taken place.}
for the qMD are taken from the hadron-string transport model UrQMD
\cite{Bass:1998ca,Bleicher:1999xi}: After the two incoming nuclei have passed through each other, (pre-)hadrons from the string and hadron dynamics of the UrQMD model
are decomposed into quarks with current masses $m_u=m_d=10$~MeV and
$m_s=150$~MeV. At the highest RHIC energy, this happens at a
center of mass time of $t=0.15$~fm/c. The quarks are then let to
evolve and interact within the qMD via a linear potential
$V(|\bold{r}_i-\bold{r}_j|)=\kappa |\bold{r}_i-\bold{r}_j|$, where $\kappa$ is the string 
tension and $\bold{r}_n$ is the position of particle $n$. Therefore the full 
Hamiltonian of the model reads:
\begin{equation} H= \sum_{i=1}^N \sqrt{ p_i^2 + m_i^2} + \frac{1}{2}
\sum_{i,j} C_{ij} V(|\bold{r}_i-\bold{r}_j|)\quad.
\label{eq:hamiltonian}
\end{equation}
Where $N$ counts the number of particles in the system and the term
$C_{ij}$ takes into account the color dependence of the interaction.

The quark--(anti-)quark interaction within this potential naturally leads to confinement through the binding of
(anti-)quarks into color neutral clusters. New hadrons are formed from quarks
whose momentum and position are close to each others. Typical values for the relative momenta of the quarks 
in the two-particle rest frame at hadronization are $|p_{q}|=|p_{\bar q}|\leq 500$~MeV, the typical distance 
is below 1~fm. Hadronization thus occurs locally into hadronic clusters of mesonic and baryonic type that 
resemble the Yo-Yo states of the LUND model. These clusters are allowed to decay in the further 
evolution of the system and the hadronization process therefore allows to conserve entropy.

As the initial state of the system is color neutral, all the 
quarks of the system will eventually gather into color neutral clusters. 
Electric charge and strangeness are conserved during the whole evolution of the system, i.e.
$s-\overline s=0$ and the net-charge equals the initial charge of the incoming nuclei.
The reader is referred to \cite{Hofmann:1999jx,Scherer:2001ap,Scherer:2005sr} for a
detailed discussion of the qMD model. Note that in the
present calculations, only $u$, $d$ and $s$ quarks are
included. Furthermore, all parton production occurs in the early stage
of the reaction during the UrQMD evolution. There is presently no
mechanism to create new (di-)quark pairs during the qMD evolution stage.
Thus, the present model provides an explicit recombination transition from quark matter to
hadronic matter in a dynamical and expanding medium. 

\begin{figure}
\includegraphics[width=0.5\textwidth]{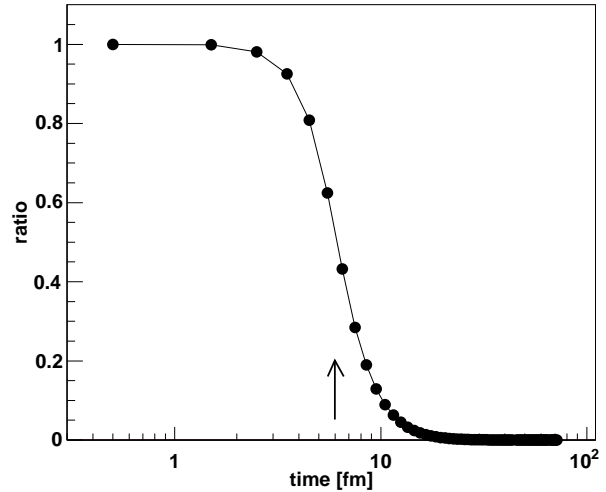}% Here is how to import EPS art
\caption{\label{fig:ParticleNumber} Fraction of the number of quarks from the total number of 
particles as a function of time at midrapidity for Au+Au reactions at $\sqrt{s_{NN}}=200$~GeV. 
The arrow depicts the hadronization time.}
\end{figure}

Let us set the stage by exploring  the time evolution of the 
hadronization dynamics in the model. Fig.~\ref{fig:ParticleNumber} depicts 
the fraction of quark matter on the total number of particles in the system 
(i.e. quark fraction $=(n_q+n_{\overline q})/(n_{\rm hadron}+n_q+n_{\overline q})$) 
as a function of time. 
One observes that the fireball stays in a deconfined state during the first
6~fm/c where almost no quarks hadronize. As the system expands further 
and the density decreases, quark recombination into baryons and mesons occurs and
the number of deconfined quarks drops to zero.
Next we turn to the investigation of the various fluctuation signals.
The electric charge ratio fluctuations were proposed as a clear signal for
the onset of the quark-gluon plasma phase \cite{Jeon:2000wg}. The
basis for the argument is that the quanta of the electric charge are
smaller in a quark gluon plasma phase than in a hadron gas and are
distributed over a larger number of particles. Moving one
charged particle from/to the rapidity window then leads to larger fluctuations in a hadron gas
than in a QGP. The electric charge ratio fluctuation can be quantified by the
measure $\tilde{D}$ defined as:
\begin{equation}
\tilde{D} = \frac{1}{C_{\mu}C_{y}} \langle N_{ch} \rangle \langle \delta R^2 \rangle_{\Delta y}\quad.
\end{equation}
Where $N_{ch}$ stands for the number of charged particles,
$R=(1+F)/(1-F)$ with $F=Q/N_{ch}$, $Q$ being the electric
charge. Following \cite{Bleicher:2000ek}, charge fluctuations are
corrected with the factors $C_{\mu}$ and $C_{y}$ to take into account
the finite acceptance. As suggested in \cite{Jeon:2000wg,Bleicher:2000ek}, the quantity $\tilde{D}$ 
is calculated in a rapidity window of $y = \pm 0.5$.
It was argued that depending on the initial nature of the system, $\tilde{D}$
will yield distinctly different results: $\tilde{D}=1$ for a quark-gluon
plasma, $\tilde{D}=2.8$ for a resonance gas and $\tilde{D}=4$ for an
uncorrelated pion gas.

Experimentally, charge ratio fluctuations have been measured at RHIC energies by STAR
\cite{Westfall:2004xy,Pruneau:2003ky} and PHENIX
\cite{Nystrand:2002pc,Adcox:2002mm}.  Both experimental analyses yield
result compatible with a hadron gas - in strong contrast to the first
expectations. Further results  from the CERN-SPS
\cite{Alt:2004ir,Sako:2004pw} based on a slightly different measure
for the charge ratio fluctuations did also  yield results compatible with
the hadronic expectation.
\begin{figure}
\includegraphics[width=0.5\textwidth]{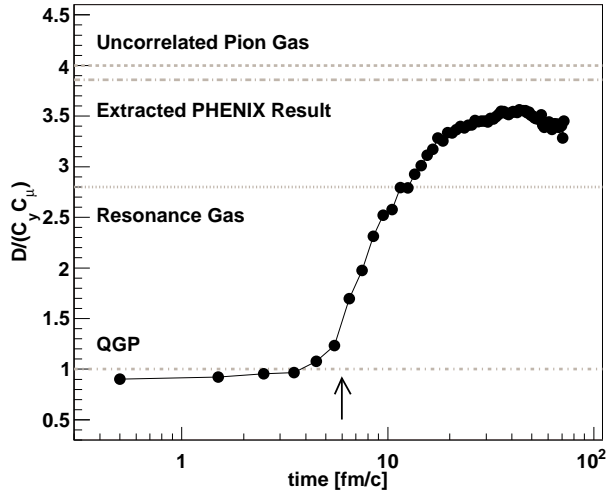}% Here is how to import EPS art
\caption{\label{fig:Dfluctuations} Corrected charge fluctuations $\tilde{D}$ as a 
function of time within the qMD model for Au+Au reaction at $\sqrt{s_{NN}}=200$~GeV (full symbols). 
Also shown are the values for an uncorrelated pion gas, a resonance gas and a quark-gluon plasma. 
The arrow depicts the hadronization time.}
\end{figure}
Fig. \ref{fig:Dfluctuations} shows the result for $\tilde{D}$ from the qMD recombination approach
as a function of time. In the early stage, when the system is completely in the deconfined phase, 
$\tilde{D}=1$ as expected. When approaching the hadronization time, $\tilde{D}$ starts to 
increase and reaches $\tilde{D} \approx 3.5$ after hadronization. As can be seen from
Fig.~\ref{fig:ParticleNumber} the increase of $\tilde{D}$ occurs exactly at
the same time as the recombination of the quarks and anti-quarks to hadrons proceeds. 
The slight decrease of $\tilde{D}$ at later times is related to the decay of resonances.

As a next observable, we now turn to charge transfer fluctuations that were also suggested to provide 
insight about the formation of a QGP phase. Charge transfer fluctuations are a  measure of the
local charge correlation length. They are  defined as \cite{Shi:2005rc,Jeon:2005kj}:
\begin{equation}
D_{u}(\eta)= \langle u(\eta)^2 \rangle - \langle u(\eta) \rangle^2\quad,
\end{equation}
with the charge transfer $u(\eta)$ being the forward-backward charge
difference:
\begin{equation}
u(\eta)=[Q_F(\eta)-Q_B(\eta)]/2\quad,
\end{equation}
where $Q_F$ and $Q_B$ are the charges in the forward and backward
hemisphere of the region separated at $\eta=0$. In our calculations, we
take a total window of $y = \pm 1$, corresponding to the STAR
acceptance. Experimental data on this oberservable is not available up to now.

Because the measured quantity is local, it can give information about
the presence and the extent of a QGP in rapidity space. Thus, one
expects to observe the lowest value of the charge transfer fluctuations
at midrapidity, where the energy density is the highest and where the
plasma is located. The local charge fluctuation is expected to be much
lower in a quark-gluon plasma than in a hadron gas.
\begin{figure}
\includegraphics[width=0.5\textwidth]{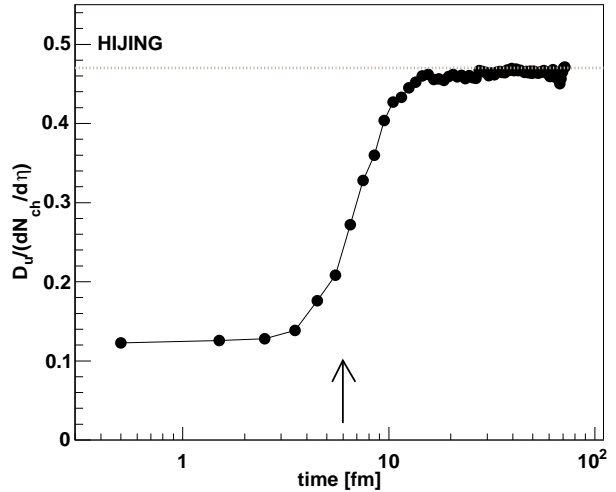}% Here is how to import EPS art
\caption{\label{fig:ChTransferFluctuations} Charge transfer fluctuations at midrapidity and $\eta=0$ as a function of 
time within the qMD model (full symbols) for Au+Au reactions at $\sqrt{s_{NN}}=200$~GeV. 
The arrow indicates the hadronization time.}
\end{figure}
The results from the present calculations are shown in Fig.~\ref{fig:ChTransferFluctuations}. 
As expected, the correlation length (at central rapidities) is small, with $D_u/(dN_{ch}/dy) \approx 0.1$, as 
long as the system is in the quark phase. However, similar to the charge ratio fluctuation discussed above,
the charge transfer measure increases with time up to its
hadronic value of $D_u/(dN_{ch}/dy) \approx 0.5$ when hadronization has happened. 
The final state result is in agreement with the value given by HIJING
calculations and therefore in line with the hadronic expectation \cite{Jeon:2005yi}.

Finally, we analyse the baryon-strangeness correlation $C_{BS}$ \cite{Koch:2005vg}. This correlation 
was proposed as a tool to study the property of the matter created in
heavy ion collisions. The baryon-strangeness correlation is defined as:
\begin{figure}
\includegraphics[width=0.5\textwidth]{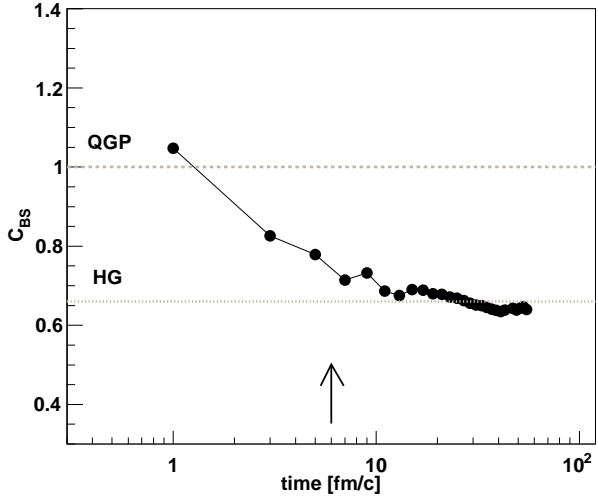}% Here is how to import EPS art
\caption{\label{fig:Cbs} $C_{BS}$ correlation coefficient as a function of time for central Au+Au 
reactions at $\sqrt{s_{NN}}=200$~GeV (full circles).  Also shown are the predicted values for a 
hadron gas and a QGP. The arrow depicts the hadronization time.}
\end{figure}
\begin{equation}
C_{BS}=-3 \frac{\langle B S \rangle - \langle B \rangle \langle S
\rangle}{ \langle S^2 \rangle - \langle S \rangle^2 }\quad,
\end{equation}
where $B$ and $S$ are the baryon number and strangeness in a
given event\footnote{In Ref. \cite{Haussler:2006mq} it was shown that the true value of $C_{BS}$ can 
be reasonably approximated  by taking into account charged kaons and $\Lambda's$ only. This might be easier for
current day experiments because it avoids the measurement of neutral particles.}. 

The rationale behind this quantity is the fact that baryon number and strangeness
are differently correlated, depending on the phase the system is
in. In an ideal weakly coupled quark-gluon plasma, strangeness will be
carried by strange quarks and is therefore strictly coupled to baryon charge. 
Thus, a clear correlation between baryon charge and strangeness is expected in a quark-gluon plasma.
The expected numerical value for an ideal QGP is $C_{BS}=1$ \cite{Koch:2005vg,Majumder:2005ai}. 
In a hadron gas on the contrary, strangeness can be carried without baryon number (e.g. with strange
mesons). As a result the correlation between strangeness and baryon number will be weakened compared 
to the quark matter scenario. The numerical value for a non-interacting hadron gas 
is $C_{BS}=0.66$ \cite{Koch:2005vg,Majumder:2005ai}.

The behaviour of $C_{BS}$ as a function of time for the dynamical recombination model under study 
is depicted in Fig.~\ref{fig:Cbs}. For early times, $C_{BS}$ starts from the expected value of unity 
in agreement with the ideal weakly coupled quark-gluon plasma value. During the course of the 
recombination of the quarks, $C_{BS}$ approaches the
hadron-gas value $C_{BS} \approx 0.6$. 

In conclusion, we have studied a variety of suggested event-by-event signatures for the
formation of a deconfined QGP state within a dynamical quark recombination approach.

The analyses was done for central Au+Au events at
$\sqrt{s_{NN}}=200$ GeV and involved charge ratio fluctuations, charge transfer fluctuations 
and baryon-strangeness correlations. For all these predicted  "smoking gun" QGP observables, 
we find that the hadronization by recombination leads to results expected for a hadron gas in the final
state. This is especially remarkable, as the initial values for these observables were identical to the
predicted QGP values. 

For all these quantities, the change of the observables from there QGP value to the hadronic gas value 
can be traced back to the recombination hadronization mechanism because the change of the quantitative values
of $\tilde D$, $D_u$ and $C_{BS}$ takes place during the time of hadronization. 

From these observations we draw two mutually converse conclusions:

\begin{enumerate}
\item
The influence of the recombination/hadronization on fluctuation probes is strong enough 
to blur the initially present QGP signature. This might explain why fluctuation 
measurements have not provided the expected proof for the
formation of a plasma of quarks and gluons. 

\item
However, if one assumes that a QGP was indeed formed at RHIC energies,
the experimental fact that all of the discussed fluctuation probes turn out to yield the hadronic value 
can be seen as a strong argument supporting recombination as the mechanism
responsible for hadronization at RHIC.
\end{enumerate}

\begin{acknowledgments}
We thanks Drs. S. A. Bass, S. Jeon, V. Koch for valuable suggestions and
comments. The computational resources have been provided by the Center
for Scientific Computing at Frankfurt. S.H. thanks the Frankfurt
International Graduate School for Science (FIGSS) for financial
support. This work was supported by GSI and BMBF.
\end{acknowledgments}


\begin{thebibliography}{99}


\bibitem{Gazdzicki:1992ri}
  M.~Gazdzicki and S.~Mrowczynski,
  %``A Method to study 'equilibration' in nucleus-nucleus collisions,''
  Z.\ Phys.\  C {\bf 54}, 127 (1992).
  %%CITATION = ZEPYA,C54,127;%%

\bibitem{Mrowczynski:1997kz}
  S.~Mrowczynski,
  %``Hadronic matter compressibility from event-by-event analysis of  heavy-ion
  %collisions,''
  Phys.\ Lett.\  B {\bf 430}, 9 (1998)
  [arXiv:nucl-th/9712030].
  %%CITATION = PHLTA,B430,9;%%

\bibitem{Bleicher:1998wd}
  M.~Bleicher {\it et al.},
  %``Fluctuations and inhomogenities of energy density and isospin in  Pb + Pb
  %at the SPS,''
  Nucl.\ Phys.\  A {\bf 638}, 391 (1998).
  %%CITATION = NUPHA,A638,391;%%

\bibitem{Bleicher:1998wu}
  M.~Bleicher {\it et al.},
  %``Can momentum correlations prove kinetic equilibration in heavy ion
  %collisions at 160-A-GeV?,''
  Phys.\ Lett.\  B {\bf 435}, 9 (1998)
  [arXiv:hep-ph/9803345].
  %%CITATION = PHLTA,B435,9;%%

\bibitem{Stephanov:1998dy}
  M.~A.~Stephanov, K.~Rajagopal and E.~V.~Shuryak,
  %``Signatures of the tricritical point in {QCD},''
  Phys.\ Rev.\ Lett.\  {\bf 81}, 4816 (1998)
  [arXiv:hep-ph/9806219].
  %%CITATION = PRLTA,81,4816;%%

\bibitem{Jeon:1999gr}
  S.~Jeon and V.~Koch,
  %``Fluctuations of particle ratios and the abundance of hadronic
  %resonances,''
  Phys.\ Rev.\ Lett.\  {\bf 83}, 5435 (1999)
  [arXiv:nucl-th/9906074].
  %%CITATION = PRLTA,83,5435;%%

\bibitem{Stephanov:1999zu}
  M.~A.~Stephanov, K.~Rajagopal and E.~V.~Shuryak,
  %``Event-by-event fluctuations in heavy ion collisions and the {QCD}  critical
  %point,''
  Phys.\ Rev.\  D {\bf 60}, 114028 (1999)
  [arXiv:hep-ph/9903292].
  %%CITATION = PHRVA,D60,114028;%%

\bibitem{Mrowczynski:1999sf}
  S.~Mrowczynski,
  %``Chemical fluctuations in high-energy nuclear collisions,''
  Phys.\ Lett.\  B {\bf 459}, 13 (1999)
  [arXiv:nucl-th/9901078].
  %%CITATION = PHLTA,B459,13;%%

\bibitem{Capella:1999uc}
  A.~Capella, E.~G.~Ferreiro and A.~B.~Kaidalov,
  %``Event-by-event fluctuations in heavy-ion collisions and the quark-gluon
  %string model,''
  Eur.\ Phys.\ J.\  C {\bf 11}, 163 (1999)
  [arXiv:hep-ph/9903338].
  %%CITATION = EPHJA,C11,163;%%

\bibitem{Mrowczynski:1999un}
  S.~Mrowczynski,
  %``Generalizing Phi-measure of event-by-event fluctuations in high-energy
  %heavy-ion collisions,''
  Phys.\ Lett.\  B {\bf 465}, 8 (1999)
  [arXiv:nucl-th/9905021].
  %%CITATION = PHLTA,B465,8;%%

\bibitem{Bleicher:2000tr}
  M.~Bleicher, J.~Randrup, R.~Snellings and X.~N.~Wang,
  %``Enhanced event-by-event fluctuations in pion multiplicity as a signal  of
  %disoriented chiral condensates at RHIC,''
  Phys.\ Rev.\  C {\bf 62}, 041901 (2000)
  [arXiv:nucl-th/0006047].
  %%CITATION = PHRVA,C62,041901;%%

\bibitem{Asakawa:2000wh}
  M.~Asakawa, U.~W.~Heinz and B.~Muller,
  %``Fluctuation probes of quark deconfinement,''
  Phys.\ Rev.\ Lett.\  {\bf 85}, 2072 (2000)
  [arXiv:hep-ph/0003169].
  %%CITATION = PRLTA,85,2072;%%

\bibitem{Bleicher:2000ek}
  M.~Bleicher, S.~Jeon and V.~Koch,
  %``Event-by-event fluctuations of the charged particle ratio from
  %non-equilibrium transport theory,''
  Phys.\ Rev.\  C {\bf 62}, 061902 (2000)
  [arXiv:hep-ph/0006201].
  %%CITATION = PHRVA,C62,061902;%%

\bibitem{Jeon:2001ka}
  S.~Jeon, V.~Koch, K.~Redlich and X.~N.~Wang,
  %``Fluctuations of rare particles as a measure of chemical equilibration,''
  Nucl.\ Phys.\  A {\bf 697}, 546 (2002)
  [arXiv:nucl-th/0105035].
  %%CITATION = NUPHA,A697,546;%%

\bibitem{Sa:2001ma}
  B.~H.~Sa, X.~Cai, A.~Tai and D.~M.~Zhou,
  %``Behavior of event-by-event fluctuations in the charged particle ratio  in
  %relativistic nucleus nucleus collisions,''
  Phys.\ Rev.\  C {\bf 66}, 044902 (2002)
  [arXiv:nucl-th/0112038].
  %%CITATION = PHRVA,C66,044902;%%

\bibitem{Koch:2001zn}
  V.~Koch, M.~Bleicher and S.~Jeon,
  %``Event-by-event fluctuations and the QGP,''
  Nucl.\ Phys.\  A {\bf 698}, 261 (2002)
  [Nucl.\ Phys.\  A {\bf 702}, 291 (2002)]
  [arXiv:nucl-th/0103084].
  %%CITATION = NUPHA,A702,291;%%

\bibitem{Hatta:2003wn}
  Y.~Hatta and M.~A.~Stephanov,
  %``Proton number fluctuation as a signal of the QCD critical end-point,''
  Phys.\ Rev.\ Lett.\  {\bf 91}, 102003 (2003)
  [Erratum-ibid.\  {\bf 91}, 129901 (2003)]
  [arXiv:hep-ph/0302002].
  %%CITATION = PRLTA,91,102003;%%

\bibitem{Ferreiro:2003dw}
  E.~G.~Ferreiro, F.~del Moral and C.~Pajares,
  %``Transverse momentum fluctuations and percolation of strings,''
  Phys.\ Rev.\  C {\bf 69}, 034901 (2004)
  [arXiv:hep-ph/0303137].
  %%CITATION = PHRVA,C69,034901;%%

\bibitem{Mekjian:2004qf}
  A.~Z.~Mekjian,
  %``Fluctuations in the statistical model of relativistic heavy ion
  %collisions,''
  Nucl.\ Phys.\  A {\bf 761}, 132 (2005)
  [arXiv:nucl-th/0411063].
  %%CITATION = NUPHA,A761,132;%%

\bibitem{Mrowczynski:2004cg}
  S.~Mrowczynski, M.~Rybczynski and Z.~Wlodarczyk,
  %``Transverse momentum versus multiplicity fluctuations in high-energy
  %nuclear collisions,''
  Phys.\ Rev.\  C {\bf 70}, 054906 (2004)
  [arXiv:nucl-th/0407012].
  %%CITATION = PHRVA,C70,054906;%%

\bibitem{Shi:2005rc}
  L.~j.~Shi and S.~Jeon,
  %``Charge transfer fluctuations as a signal for QGP,''
  Phys.\ Rev.\  C {\bf 72}, 034904 (2005)
  [arXiv:hep-ph/0503085].
  %%CITATION = PHRVA,C72,034904;%%

\bibitem{Jeon:2005kj}
  S.~Jeon, L.~Shi and M.~Bleicher,
  %``Detecting QGP with charge transfer fluctuations,''
  Phys.\ Rev.\  C {\bf 73}, 014905 (2006)
  [arXiv:nucl-th/0506025].
  %%CITATION = PHRVA,C73,014905;%%

\bibitem{Koch:2005vg}
  V.~Koch, A.~Majumder and J.~Randrup,
  %``Baryon-strangeness correlations: A diagnostic of strongly interacting
  %matter,''
  Phys.\ Rev.\ Lett.\  {\bf 95}, 182301 (2005)
  [arXiv:nucl-th/0505052].
  %%CITATION = PRLTA,95,182301;%%

\bibitem{Konchakovski:2005hq}
  V.~P.~Konchakovski, S.~Haussler, M.~I.~Gorenstein, E.~L.~Bratkovskaya, M.~Bleicher and H.~Stoecker,
  %``Particle number fluctuations in high energy nucleus nucleus collisions
  %from microscopic transport approaches,''
  Phys.\ Rev.\  C {\bf 73}, 034902 (2006)
  [arXiv:nucl-th/0511083].
  %%CITATION = PHRVA,C73,034902;%%

\bibitem{Cunqueiro:2005hx}
  L.~Cunqueiro, E.~G.~Ferreiro, F.~del Moral and C.~Pajares,
  %``Multiplicity fluctuations in the string clustering approach,''
  Phys.\ Rev.\  C {\bf 72}, 024907 (2005)
  [arXiv:hep-ph/0505197].
  %%CITATION = PHRVA,C72,024907;%%

\bibitem{Torrieri:2005va}
  G.~Torrieri, S.~Jeon and J.~Rafelski,
  %``Particle yield fluctuations and chemical non-equilibrium at RHIC,''
  Phys.\ Rev.\  C {\bf 74}, 024901 (2006)
  [arXiv:nucl-th/0503026].
  %%CITATION = PHRVA,C74,024901;%%

\bibitem{Armesto:2006bv}
  N.~Armesto, L.~McLerran and C.~Pajares,
  %``Long range forward-backward correlations and the color glass condensate,''
  Nucl.\ Phys.\  A {\bf 781}, 201 (2007)
  [arXiv:hep-ph/0607345].
  %%CITATION = NUPHA,A781,201;%%

\bibitem{Jeon:2003gk}
  S.~Jeon and V.~Koch,
  ``Event-by-event fluctuations,''
in Quark-Gluon Plasma 3, eds. R.C. Hwa and X.N Wang, World Scientific Singapore, p. 430-490
  [arXiv:hep-ph/0304012].
  %%CITATION = HEP-PH/0304012;%%

\bibitem{Jeon:2000wg}
  S.~Jeon and V.~Koch,
  %``Charged particle ratio fluctuation as a signal for QGP,''
  Phys.\ Rev.\ Lett.\  {\bf 85}, 2076 (2000)
  [arXiv:hep-ph/0003168].
  %%CITATION = PRLTA,85,2076;%%

\bibitem{Jeon:2001ue}
  S.~Jeon and S.~Pratt,
  %``Balance functions, correlations, charge fluctuations and  interferometry,''
  Phys.\ Rev.\  C {\bf 65}, 044902 (2002)
  [arXiv:hep-ph/0110043].
  %%CITATION = PHRVA,C65,044902;%%

\bibitem{Zhang:2002dy}
  Q.~H.~Zhang, V.~Topor Pop, S.~Jeon and C.~Gale,
  %``Charged particle ratio fluctuations and microscopic models of nuclear
  %collisions,''
  Phys.\ Rev.\  C {\bf 66}, 014909 (2002)
  [arXiv:hep-ph/0202057].
  %%CITATION = PHRVA,C66,014909;%%

\bibitem{Pruneau:2002yf}
  C.~Pruneau, S.~Gavin and S.~Voloshin,
  %``Methods for the study of particle production fluctuations,''
  Phys.\ Rev.\  C {\bf 66}, 044904 (2002)
  [arXiv:nucl-ex/0204011].
  %%CITATION = PHRVA,C66,044904;%%

\bibitem{Haussler:2005ei}
  S.~Haussler, H.~Stoecker and M.~Bleicher,
  %``Event-by-event analysis of baryon-strangeness correlations: Pinning  down
  %the critical temperature and volume of QGP formation,''
  Phys.\ Rev.\  C {\bf 73}, 021901 (2006)
  [arXiv:hep-ph/0507189].
  %%CITATION = PHRVA,C73,021901;%%

\bibitem{Shuryak:2000pd}
  E.~V.~Shuryak and M.~A.~Stephanov,
  %``When can long range charge fluctuations serve as a QGP signal?,''
  Phys.\ Rev.\  C {\bf 63}, 064903 (2001)
  [arXiv:hep-ph/0010100].
  %%CITATION = PHRVA,C63,064903;%%

\bibitem{Molnar:2003ff}
  D.~Molnar and S.~A.~Voloshin,
  %``Elliptic flow at large transverse momenta from quark coalescence,''
  Phys.\ Rev.\ Lett.\  {\bf 91}, 092301 (2003)
  [arXiv:nucl-th/0302014].
  %%CITATION = PRLTA,91,092301;%%

\bibitem{Hwa:2003ic}
  R.~C.~Hwa and C.~B.~Yang,
  %``Recombination model for fragmentation: Parton shower distributions,''
  Phys.\ Rev.\  C {\bf 70}, 024904 (2004)
  [arXiv:hep-ph/0312271].
  %%CITATION = PHRVA,C70,024904;%%

\bibitem{Fries:2003vb}
  R.~J.~Fries, B.~Muller, C.~Nonaka and S.~A.~Bass,
  %``Hadronization in heavy ion collisions: Recombination and fragmentation  of
  %partons,''
  Phys.\ Rev.\ Lett.\  {\bf 90}, 202303 (2003)
  [arXiv:nucl-th/0301087].
  %%CITATION = PRLTA,90,202303;%%

\bibitem{Fries:2003kq}
  R.~J.~Fries, B.~Muller, C.~Nonaka and S.~A.~Bass,
  %``Hadron production in heavy ion collisions: Fragmentation and  recombination
  %from a dense parton phase,''
  Phys.\ Rev.\  C {\bf 68}, 044902 (2003)
  [arXiv:nucl-th/0306027].
  %%CITATION = PHRVA,C68,044902;%%

\bibitem{Greco:2003xt}
  V.~Greco, C.~M.~Ko and P.~Levai,
  %``Parton coalescence and antiproton/pion anomaly at RHIC,''
  Phys.\ Rev.\ Lett.\  {\bf 90}, 202302 (2003)
  [arXiv:nucl-th/0301093].
  %%CITATION = PRLTA,90,202302;%%

\bibitem{Adams:2003am}
  J.~Adams {\it et al.}  [STAR Collaboration],
  %``Particle dependence of azimuthal anisotropy and nuclear modification of
  %particle production at moderate p(T) in Au + Au collisions at  s(NN)**(1/2) =
  %200-GeV,''
  Phys.\ Rev.\ Lett.\  {\bf 92}, 052302 (2004)
  [arXiv:nucl-ex/0306007].
  %%CITATION = PRLTA,92,052302;%%

\bibitem{Adler:2003kt}
  S.~S.~Adler {\it et al.}  [PHENIX Collaboration],
  %``Elliptic flow of identified hadrons in Au + Au collisions at  s(NN)**(1/2)
  %= 200-GeV,''
  Phys.\ Rev.\ Lett.\  {\bf 91}, 182301 (2003)
  [arXiv:nucl-ex/0305013].
  %%CITATION = PRLTA,91,182301;%%

\bibitem{Nonaka:2005vr}
  C.~Nonaka, B.~Muller, S.~A.~Bass and M.~Asakawa,
  %``Possible resolutions of the D-paradox,''
  Phys.\ Rev.\  C {\bf 71}, 051901 (2005)
  [arXiv:nucl-th/0501028].
  %%CITATION = PHRVA,C71,051901;%%

\bibitem{Hofmann:1999jx}
  M.~Hofmann, M.~Bleicher, S.~Scherer, L.~Neise, H.~Stoecker and W.~Greiner,
  %``Statistical mechanics of semi-classical colored objects,''
  Phys.\ Lett.\  B {\bf 478}, 161 (2000)
  [arXiv:nucl-th/9908030].
  %%CITATION = PHLTA,B478,161;%%

\bibitem{Scherer:2001ap}
  S.~Scherer, M.~Hofmann, M.~Bleicher, L.~Neise, H.~Stoecker and W.~Greiner,
  %``Microscopic coloured quark-dynamics in the soft non-perturbative  regime:
  %Description of hadron formation in relativistic S + Au  collisions at CERN,''
  New J.\ Phys.\  {\bf 3}, 8 (2001)
  [arXiv:nucl-th/0106036].
  %%CITATION = NJOPF,3,8;%%

\bibitem{Scherer:2005sr}
  S.~Scherer and H.~Stoecker,
  %``Multifragmentation, clustering, and coalescence in nuclear collisions,''
  arXiv:nucl-th/0502069.
  %%CITATION = NUCL-TH/0502069;%%

\bibitem{Hofmann:1999jy}
  M.~Hofmann, J.~M.~Eisenberg, S.~Scherer, M.~Bleicher, L.~Neise, H.~Stoecker and W.~Greiner,
  %``Nonequilibrium dynamics of a hadronizing quark-gluon plasma,''
  arXiv:nucl-th/9908031.
  %%CITATION = NUCL-TH/9908031;%%

\bibitem{Bass:1998ca}
  S.~A.~Bass {\it et al.},
  %``Microscopic models for ultrarelativistic heavy ion collisions,''
  Prog.\ Part.\ Nucl.\ Phys.\  {\bf 41}, 225 (1998)
  [arXiv:nucl-th/9803035].
  %%CITATION = PPNPD,41,225;%%

\bibitem{Bleicher:1999xi}
  M.~Bleicher {\it et al.},
  %``Relativistic hadron hadron collisions in the ultra-relativistic quantum
  %molecular dynamics model,''
  J.\ Phys.\ G {\bf 25}, 1859 (1999)
  [arXiv:hep-ph/9909407].
  %%CITATION = JPHGB,G25,1859;%%

\bibitem{Westfall:2004xy}
  G.~D.~Westfall  [STAR collaboration],
  %``Correlations and fluctuations in STAR,''
  J.\ Phys.\ G {\bf 30}, S1389 (2004)
  [arXiv:nucl-ex/0404004].
  %%CITATION = JPHGB,G30,S1389;%%

\bibitem{Pruneau:2003ky}
  C.~A.~Pruneau  [STAR Collaboration],
  %``Event by event net charge fluctuations,''
  Heavy Ion Phys.\  {\bf 21}, 261 (2004)
  [arXiv:nucl-ex/0304021].
  %%CITATION = APHPF,21,261;%%

\bibitem{Nystrand:2002pc}
  J.~Nystrand  [PHENIX Collaboration],
  %``Charge fluctuations at mid-rapidity in Au + Au collisions in the PHENIX
  %experiment at RHIC,''
  Nucl.\ Phys.\  A {\bf 715}, 603 (2003)
  [arXiv:nucl-ex/0209019].
  %%CITATION = NUPHA,A715,603;%%

\bibitem{Adcox:2002mm}
  K.~Adcox {\it et al.}  [PHENIX Collaboration],
  %``Net charge fluctuations in Au + Au interactions at s(NN)**(1/2) =
  %130-GeV,''
  Phys.\ Rev.\ Lett.\  {\bf 89}, 082301 (2002)
  [arXiv:nucl-ex/0203014].
  %%CITATION = PRLTA,89,082301;%%

\bibitem{Alt:2004ir}
  C.~Alt {\it et al.}  [NA49 Collaboration],
  %``Electric charge fluctuations in central Pb + Pb collisions at 20-AGeV,
  %30-AGeV, 40-AGeV, 80-AGeV and 158-AGeV,''
  Phys.\ Rev.\  C {\bf 70}, 064903 (2004)
  [arXiv:nucl-ex/0406013].
  %%CITATION = PHRVA,C70,064903;%%

\bibitem{Sako:2004pw}
  H.~Sako and H.~Appelshaeuser  [CERES/NA45 Collaboration],
  %``Event-by-event fluctuations at 40-A-GeV/c, 80-A-GeV/c, and 158-A-GeV/c  in
  %Pb + Au collisions,''
  J.\ Phys.\ G {\bf 30}, S1371 (2004)
  [arXiv:nucl-ex/0403037].
  %%CITATION = JPHGB,G30,S1371;%%

\bibitem{Jeon:2005yi}
  S.~Jeon, L.~Shi and M.~Bleicher,
  %``Charge transfer fluctuations as a QGP signal,''
  J.\ Phys.\ Conf.\ Ser.\  {\bf 27}, 194 (2005)
  [arXiv:nucl-th/0511066].
  %%CITATION = 00462,27,194;%%

\bibitem{Haussler:2006mq}
  S.~Haussler, S.~Scherer and M.~Bleicher,
  %``Baryon - strangeness correlations from hadron / string- and
  %quark-dynamics,''
  arXiv:nucl-th/0611002.
  %%CITATION = NUCL-TH/0611002;%%

\bibitem{Majumder:2005ai}
  A.~Majumder, V.~Koch and J.~Randrup,
  %``Baryon number and strangeness: Signals of a deconfined antecedent,''
  J.\ Phys.\ Conf.\ Ser.\  {\bf 27}, 184 (2005)
  [arXiv:nucl-th/0510037].
  %%CITATION = 00462,27,184;%%



\end{thebibliography}
\end{document}